\documentclass[twocolumn,pre]{revtex4-1}
\usepackage{amsmath}
\usepackage{amsfonts}

\usepackage{graphicx}
\renewcommand{\Pr}[1]{\operatorname{P}\left(#1 \right)}
\renewcommand{\P}{\mathbb{P}}
\newcommand{\avg}[1]{\left\langle #1 \right\rangle}
\newcommand{\Z}{\mathcal Z_\text{kin}}
\newcommand{\pd}[2]{\frac{\partial #1}{\partial #2}}
\newcommand{\pdel}[2]{\frac{\delta #1}{\delta #2}}

\begin{document}

\title{ Unifying Theories for Nonequilibrium Statistical Mechanics}
%\title{ Maximum Entropy Closure for Nonequilibrium Statistical Mechanics}
\author{David M. Rogers}
\affiliation{University of South Florida}

% change author names for approach names?
\begin{abstract}
  The question of deriving general force/flux relationships that apply
out of the linear response regime is a central topic of
theories for nonequilibrium statistical mechanics.
This work applies an information theory perspective to compute
approximate force/flux relations and compares the
result with traditional alternatives.
If it can be said that there is a consensus on the {\em form} of response theories in driven,
nonequilibrium transient dynamics, then that consensus is consistent
with maximizing the entropy of a distribution over transition space. %and their symmetries are essentially solved problems.
%We find that a simple maximum entropy approach highlights
%key points of agreement with the projector-operator based fluctuation-dissipation theorem,
%the fluctuation theorem, large deviation theory and the chaotic hypothesis.
This agreement requires the problem of force/flux relationships to
be described entirely in terms of such transition distributions,
rather than steady-state properties (such as near-equilibrium works) or distributions
over trajectory space (such as maximum caliber).
Within the transition space paradigm, it is actually simpler to work in the fully nonlinear regime
without relying on any assumptions about the steady-state or long-time properties.
Our results are compared to extensive numerical simulations of two
very different systems.  The first is a the periodic Lorentz gas under constant external force,
extended with angular velocity and physically realistic inelastic scattering.
The second is an $\alpha$-Fermi-Pasta-Ulam chain, extended with a Langevin
thermostat that couples only to individual harmonic modes.
Although we simulate both starting from transient initial conditions,
the maximum entropy structure of the transition distribution is clearly evident
on both atomistic and intermediate size scales.
The result encourages further development of empirical laws for nonequilibrium
statistical mechanics by employing analogies with standard
maximum entropy techniques -- even in cases where large deviation principles
cannot be rigorously proven.
%Outstanding areas of debate between the approaches, related to
%existence and use of steady-states, how to define appropriate
%coarse-grained coordinates, and how to interpret the dissipation function,
%do not alter the main conclusion.
\end{abstract}

% 05.40.-a Fluctuation phenomena, random processes, noise, and Brownian motion
% 05.45.Pq	Numerical simulations of chaotic systems
% 05.70.Ln	Nonequilibrium and irreversible thermodynamics
%\keywords{} 3--7 required!

\maketitle{}

\section{ Introduction}

  There is a growing consensus\cite{cbust05,cmaes08,cjarz15,rchet15} that theories
describing the full force/flux curve are connected by simple, general principles.
However, the most compelling, simple examples are based
on proving large deviation laws for sums of random numbers,\cite{svara08} empirical
distributions\cite{fbone97} or Markov chains.\cite{htouc09}
In this mathematical context, it can be difficult to make creative applications to
simple physical systems, like a rotating dipole or fluid flow through a channel.
Our goal in this work is to present an alternative point of view
on the large deviation theory by showing how it is implied by chosing
a canonical, maximum entropy, form for the statistical mechanics
of force/flux relations in nonequilibrium dynamics.
%that was workable in other approaches, but has been historically under-appreciated.

  There are multiple theories of nonequilibrium statistical mechanics that have developed
into essentially complete programs for studying stochastic molecular systems
driven by external forces.  Perhaps the earliest among these is thermodynamics
itself, originally developed to describe the energy flows in engines driven by
nonequilibrium flows of work and heat.  The first and second laws are founded on the laws
of conservation of energy, volume, mass, and charge, and therefore apply to
all macroscopic nonequilibrium situations.  Moreover,
the equilibrium relations provide a default model
for reservoirs that store and deliver these quantities from the laboratory
into an arbitrary dynamical system in any state.\cite{droge12}
%to reservoirs at thermodynamic equilibrium leads to the well-known
%definitions of heat, work, and the inequalities involving entropy.
In the thermodynamic limit, the equilibrium theory of statistical mechanics
predicts the general form for probabilities of conserved quantities
from information about the environmental reservoirs.\cite{ejayn57}

  It is the goal of nonequilibrium statistical mechanics to provide the general form for
rates of movement of conserved quantities within and between systems.  Such equations of
motion are the nonequilibrium analogues for the equilibrium equations of state.  Also
known as force/flux relationships, these equations of motion should give
probabilities for the kinetics of processes given information
about the state of the system and environment.

  The peculiar approach that will be taken in this work is to
tackle the subjective problem of ascribing probabilities to the motion of
a physical system that is interacting with a noisy environment.
Since the environment will only be described in a statistical sense,
the resulting probabilities may be greatly in error if there are conserved
quantities in the dynamics that are not accounted for in the model.
This is exactly the old problem with assuming the ergodic hypothesis
when applying equilibrium statistical mechanics.\cite{jlebo73,ejayn79}
We paraphrase Jaynes in claiming a dual use for the results so obtained.
Where the results are accurate, it provides us a canonical form
for nonequilibrium thermodynamics.  Where they disagree with experiment
(either observations from physical or more accurate theoretical models),
the disagreement shows evidence that the maximum entropy procedure
did not account for relevant, reproducible information.
Failure of the `canonical nonequilibrium' model prompts us to
search for additional conserved quantities in the dynamics,
and will thus lead to new discoveries.

  The sections that follow lay out the `canonical' form
in full, and then describe its application to our two examples.
%We then give a very brief catalogue of known results for each.
%Next, the maximum entropy formalism
%is derived in a way that shows both its close connection to
%the theory of large deviations and to equilibrium statistical mechanics.
%This insight is incorporated into a third theory, where
%energy balances enter naturally into the expression for the
%transition probability.
Subsequently, we show how this `canonical' idea was implicit in
the original Mori-Zwanzig and Green-Kubo theories.
It was not generally recognized, however, because historical
applications of those theories used many specializations
appropriate only for steady-states near equilibrium.
Next, we show that it predicts a forward fluctuation
relation that is more general (but less rigorously applicable)
than standard fluctuation theorems.
Because it can apply to irreversible processes, the fluctuation
relation may prove more helpful for comparing to experiments.
%focused on predicting long-time behavior,
%which confined their use to near-equilibrium situations.

%  Differences between the theories, discussed in the conclusion,
%warrant special attention for ill-behaved systems, where poor choices
%of coarse coordinates exist or where external forces can cause
%`blow-up' of the dynamics.  Despite these open topics, it can
%be concluded that the use of maximum
%transition entropy to generate a statistical description of
%nonequilibrium transitions in terms of forces and flows
%is well enough established to become a standard technique
%from which generalized Langevin equations and continuum
%hydrodynamic equations can be derived within a sequence
%of increasingly accurate approximations.

\section{ Maximum Transition Entropy}\label{s:maxtrans}

  This section derives our `canonical form' for the transition probability distribution.
%Although this form can be used to derive the transient fluctuation theorems,
%it does not center on the operation of time-reversal.
%Instead, it is based completely on an
We start by assuming there is some probability space of possible transitions,
$g \in \mathcal G$, with an unknown underlying measure, $d\mu(g)$.
For continuous distributions, $d\mu(g)/dg \equiv P^0(g)$ is the probability distribution of $g$.
Time is discretized, $t_1 \le t_2 \le \ldots$, according to any useful convention
(equal time slices, first collision time, etc.), and each transition
(labeled $g_j$ for the transition taking $t_j \to t_{j+1})$ is associated
with some (usually bounded) flow, $J(g_j)$.  Each flow must measure
the exchange of a conserved quantity ($J$) between the system and its environment.
Here, conserved means that the flows would all be exactly
zero if the system were not interacting with an external environment.

  With this setup, we can phrase a maximum entropy problem as
follows: Find the probability measure, $d\nu$, which maximizes
the relative entropy,
\begin{equation}
S[d\nu | d\mu] = -\int_{\mathcal G} d\nu \log \frac{d\nu}{d\mu}
\end{equation}
under the constraint,
\begin{equation}
\avg{J} = \int J \, d\nu =: \avg{J}
.
\end{equation}
The solution is just the usual canonical distribution,
\begin{equation}
d\nu = d\mu e^{\lambda J} / \Z(\lambda)
,\label{e:pred}
\end{equation}
(so that $\Pr{g | \lambda} = d\nu/dg$)
with Lagrange multiplier determined by the derivative of
\begin{equation}
\Z(\lambda) = \int_{\mathcal G} e^{\lambda J} d\mu
.
\end{equation}

  The motivation for using maximum entropy here is a subjective uncertainty about
the underlying stochastic process.\cite{cmaes99,droge11a}
The ending result is the tilted exponential\cite{djian03,lrond07,lbert15}
and large deviation functions
announced and studied by several authors.\cite{ggall96,htouc09,rchet15}
However, we do not rely on complete knowledge of the underlying
`default' measure, $d\mu$.
In the results below (as in experimental tests), $d\mu$ is treated as an empirical observable.

  In the case of particle trajectories, $x(t)$, $d\nu$ can be chosen according to another
maximum entropy principle involving the uniform measure for $d\mu$.
Both of the mechanical systems studied here were coupled to stochastic
external reservoirs by defining the transition events,
$g$, to be equal to deviations from an Euler-Lagrange equation of motion,
\begin{equation}
g = \pdel{\mathcal A[x]}{x}
\end{equation}
where $\mathcal A$ is a classical action functional.  It turns out
that this ansatz has a plausible origin in quantum decoherence.\cite{droge17c,ldios18}
% which is all I really set out to prove in this career
Maximum entropy constraints are placed on
\begin{equation}
\avg{D} \equiv \avg{g^2 dt}, \text{ and } \avg{dE} \equiv -\avg{g\, dx}
.\label{e:en}
\end{equation}
The first Lagrange multiplier is arbitrary (we use $1/2\sigma^2$), but the second appears
to always be $\beta/2$, where $\beta^{-1} \equiv k_B T$ is the thermal temperature.
The result for $\pdel{\mathcal A[x]}{x} = F(x) - \dot p$ (force minus momentum change)
is the Langevin equation,\cite{droge12}
\begin{equation}
\Pr{g = F - \dot p | x, p} \propto e^{-g^2 dt/2\sigma^2 + \beta dx \cdot g/2} \label{e:langevin}
.
\end{equation}
So we see that the Langevin equation is `canonical.'

  It is clear that Eq.~\ref{e:pred} has the same, canonical, structure as
equilibrium statistical mechanics.  In particular, the forces and average
fluxes are conjugate thermodynamic variables,
\begin{align}
\avg{J | \lambda,\Gamma} &= \pd{\log \Z(\lambda,\Gamma)}{\lambda} \label{e:fdt2} \\
\avg{\delta J^2 | \lambda,\Gamma} &= \pd{^2 \log \Z(\lambda,\Gamma)}{\lambda^2} \label{e:fdt3}
\end{align}
These averages are conditional on the starting-point, $\Gamma$,
by the dependence of $\mathcal G, d\mu$ on $\Gamma$.
Ref.~\cite{ggall96} indirectly showed their use for deriving Onsager reciprocity.
The full, causal, analogue of the Green-Kubo relations was demonstrated in Ref~\citenum{droge11a}.

  New relations between transition probabilities can be shown directly from the ratio of
Eq.~\ref{e:pred} at two different applied forces,
\begin{equation}
\log \frac{\Pr{J | \lambda'}}{\Pr{J | \lambda}} =  (\lambda' - \lambda) J - \log \frac{\Z(\lambda')}{\Z(\lambda)}
\label{e:maxtrans}
\end{equation}
The relation obviously holds for the generalized Langevin equation (Eq.~\ref{e:langevin}),
and likewise whenever `canonical' nonequilibrium statistical mechanics applies.
However, because of its origin in maximum entropy rather than exact dynamics,
it is better to be named a (forward) fluctuation relation then a fluctuation theorem proper.
It should be qualified as `forward' because it does not rely on time-reversal
symmetry, but instead relies on conservation laws (associated by Noether's theorem
to continuous symmetries).

  Figure~\ref{f:ft} illustrates this maximum entropy structure by plotting the
ratios, $P(J | F') / P(J | F)$ for successive values of the applied force, $F = E$ or $\Delta \beta$.
The probabilities were calculated from histograms of the final bin number (Lorentz gas)
or total heat conduction (FPU lattice) using 102,400 independent trajectories
for each value of the forcing.
Despite the transient initial conditions, the relatively short simulation times,
and the nonlinearity in the flux-force curves,
the MaxTrans postulate, Eq.~\ref{e:pred}, appears to hold
for both systems examined here.

  The correspondence between $\lambda$ and the applied force,
($E$ or $\Delta \beta$) is not direct. %  Section~\ref{s:gaus} showed that none of the theories promised
%that it would be.   Instead, nonequilibrium theories provide starting points
%and frameworks for finding analytical solutions through integrating
%the equations of motion.  Many of these have been well-studied for the problems considered here.
%In the general case, Eq.~\ref{e:maxtrans}
%predicts only the relationship between the slope of $\log \Pr{J}$ and
%$d\avg{J}/d\lambda$.
% Nevertheless, this relationship applies almost universally to out-of-equilibrium cases
%that can be shown to have finite process memory.
Instead, MaxTrans only predicts a canonical form for transition probability distributions.
In the same spirit as the Boltzmann/Gibbs distribution, $\lambda$ and $J$ are a conjugate
pair, and their relation to a physical external field, $E$, can be described by some function, $\lambda(E)$.
This relationship between generalized forces,
$\lambda$, and an applied physical force, is identifiable by any of three equivalent
methods:
\begin{enumerate}
%{\em i})
\item checking the ratio of Eq.~\ref{e:maxtrans} as a function of $J$ for two different physical forces,
% {\em ii})
\item matching mean and variance of the flux to the expansion,
$\avg{J|\lambda'} = \avg{J|\lambda} + (\lambda' - \lambda) \avg{\delta J^2 |\lambda} + O(\Delta \lambda^2)$,
%{\em iii})
\item Green-Kubo style integration of the conjugate flux starting from $\lambda = E$
at short time-scales (compare Eq.~\ref{e:langevin} to Eq.~\ref{e:bias}),\cite{droge12,devan16} or
%{\em iv})
\item differentiating $\lambda(J) = d\sigma(\avg{J}) / d\avg{J}$, where
\begin{equation}
\sigma(\avg{J}) = -\int dJ\; \Pr{J|\lambda} \log \frac{\Pr{J|\lambda}}{\Pr{J|0}}.
\end{equation}
\end{enumerate}

% TODO: cite Maes here
% TODO: cite Pinkus here?

\section{ Dynamical Systems Investigated}
\subsection{ Inelastic Periodic Lorentz Gas}

  The periodic Lorentz gas describes a system of fixed scattering centers
that cause rigid-body collisions of a single, spherical gas particle.  The deflections
of the studied gas particle cause it to undergo a random walk, mimicking
an ideal gas.
We simulated free flight of a single particle under constant external field
($\vec E = E \hat x - g\hat z \in \mathbb R^2$) as a series of parabolic segments interrupted by discrete collisions.
Scattering centers were placed on a regular 2D hexagonal lattice
with side length $L$. 
Numerically, collisions were detected by solving the quartic equation required to find
the time of intersection of parabolic trajectories with one circular scatterer at the origin.
By monitoring collisions with the unit cell boundaries and translating appropriately,
only one particle-scatterer interaction needed testing during each computational update cycle.

  On each collision, the particle's location is unchanged,
and an impulsive force is chosen at random following
Eq.~\ref{e:langevin}.  An extra maximum-entropy constraint is added to
enforce reflection of the particle's velocity.  The geometry in Fig.~\ref{f:impulse}
is used in the following and defines the decomposition
of the particle's center of mass velocity into normal and tangential
components and shows its angular velocity.
We assume the particle is a uniform circular disk of radius $r$
with mass $M$ and moment of inertia $M r^2/2$.
The angular velocity is not considered in most treatments
of the Lorentz gas, but must be included for a consistent set of energy equations.
It is also needed to compute the tangential velocity, $v_t$, at the contact point.

  Straightforward application of Eq.~\ref{e:langevin} would lead to
\begin{align}
dp_n &= -(\lambda + \beta/2) \sigma^2 v_n dt + \sigma d\mathcal W_n \\
dp_t &= -(\beta \sigma^2/2) v_t dt + \sigma d\mathcal W_t
.
\end{align}
Here, $dp_n, dp_t$ represent the normal and tangential forces
added to the particle during the period of contact
and $d\mathcal W_n, d\mathcal W_t$ are independent Wiener processes.
To reach the impulsive force limit, we insist that an ``inelasticity parameter''
$\gamma \equiv \tilde\beta \sigma^2 dt/2 M$ remains
finite in the limit $dt \to 0$
so that $\sigma d\mathcal W = R \sqrt{2 M \gamma / \tilde \beta}$,
with $R$ a sample from the standard normal distribution.
The impulses (now labeled $I_n,I_t$) are then drawn from two standard normal distributions ($R_n, R_t$),
\begin{align}
I_n/M &= -\frac{\lambda \sigma^2}{M} v_n dt - \gamma v_n + R_n \sqrt{2 \gamma/M\tilde \beta} \\
I_t/M &= -\gamma v_t + R_t \sqrt{2 \gamma/M\tilde \beta}
\end{align}
This work used $\gamma = 0.01$. We also set $\lambda \sigma^2 dt / M = 2$
to accomplish perfect reflection when $\gamma = 0$.

  Adding this impulsive force to a rigid body results
in the following stochastic map, $v \mapsto M(v) = v'$, for updating
all velocity components,
\begin{align}
-v_n' &= (1 - \gamma) v_n + R_n \sqrt{2 \gamma/\tilde \beta M} \label{e:map} \\
v_t' - v_t &= -\gamma (v_t + r \omega) + R_t \sqrt{2 \gamma/\tilde \beta M} \\
r (\omega' - \omega)/2 &= v_t' - v_t
.
\end{align}
The particle's radius, $r$, need not be specified separately, since
the equation of motion depends only on the product, $r\omega$.
%can be absorbed into the units of the
%angular velocity because the moment of inertia of the disk is $Mr^2/2$.

  To show that $\tilde \beta = \beta$, it can be verified
that the Boltzmann distribution,
\begin{equation}
P(v_n,v_t,r\omega) \propto e^{-\frac{\beta M}{2} ((r\omega)^2/2 + v_n^2 + v_t^2)}
,\label{e:boltz}
\end{equation}
is a steady-state of the map.
This works in the limit where $\gamma << 1$.
The proof of the steady-state is most easily accomplished by
multiplying the moment generating functions of Eqns.~\ref{e:boltz}
and the Gaussian distribution implied by Eq~\ref{e:map}.
Our numerical simulations used $\tilde \beta$
that is required for arbitrary $\gamma$ to achieve
the fixed inverse temperature, $(\beta M)^{-1} = 5.292\cdot 10^{-19}~\tau^2/L^2$.
The minor difference between $\beta$ and $\tilde \beta$
is created because the impulse should occur
at the center of a timestep, as for the Stratonovich stochastic calculus.

  Although highly unlikely because of the extremely large value of $\beta M$ used here,
it is technically possible that the random increment to $v_n$ causes
$v_n'$ to remain inward.  Our simulation is therefore set to
sample the appropriate truncated Gaussian for $R_n$ by
generating random trials until one is found that leaves $v_n'$ pointing outward (away
from the scatterer).
Random noise is required by the fluctuation-dissipation theorem.
With friction but no random noise, numerical simulations showed
a few trajectories that settled into a stable limit cycle,
stuck bouncing back and forth between the same two scatterers.
The angular momentum did not play a role in the limit cycle,
since it went quickly to zero.  Our simulations included the small random
noise, eliminating such occurrences.

%normal component of the particle's velocity is reversed, and
%a Langevin impulse is applied to its velocity.  The thermostat
%is important to dissipate energy added by the constant external force.
%Because the object is macroscopic, the random impulses are negligible.
%To be consistent, the tangential velocity is subject to an independent
%impulsive Langevin scattering.  This required tracking the particle's angular
%velocity.

\begin{figure*}
\centering
\vspace{-3cm}
\includegraphics[width=0.9\textwidth]{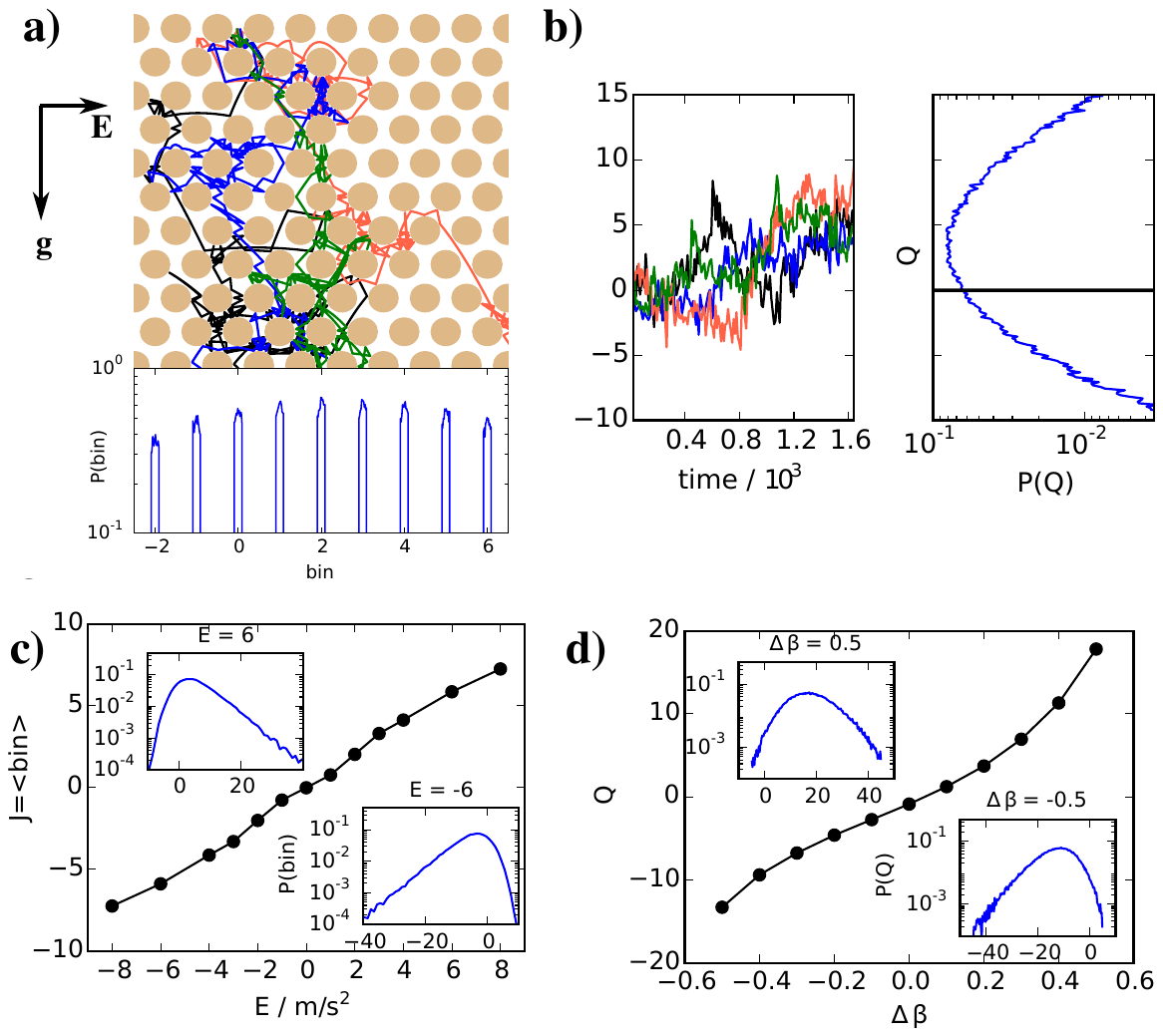}\vspace{-5.5cm}
 \caption{Tracking the horizontal flow through the Galton board setup (left panels)
or the heat flux from mode 7 to mode 3 in the Fermi-Pasta-Ulam lattice (right panels).
Example trajectories, along with a histogram of total flux at a single applied force are
shown in a,b, while c,d summarize all simulations by showing
the average flux as a function of applied force.
The insets of c,d show histograms of the flux at large values of the force.}\label{f:sys}
\end{figure*}

  Figure~\ref{f:sys}a shows four randomly chosen
trajectories for the system, along with the complete
histogram collected at row 10 for a small positive
value of $E$.  The flux, $J$, is identical to the final bin
number, determined from the $x$-coordinate.  The results presented here were
collected from the hexagonal lattice shown
with particle-to-scatter contact distance of 0.4$L$.
%, where $L$ is the lattice spacing that
%fixes the dimension of length.
The time scale, $\tau$, was set so that
the gravitation constant is $9.8 m/s^2 = 1 L/\tau^2$.
102,400 trajectories were simulated with uniform random starting locations
on the line $y=0, x\in (-0.1L,0.1L)$ and velocities chosen from
a Gaussian distribution with variance
$(\beta M)^{-1} = 5.292\cdot 10^{-19}~\tau^2/L^2$.
%$(\beta M)^{-1} = 6.5\cdot 10^{-17}$.
This is consistent with a physical
scatterer diameter of 6.35 cm and mass of 5 g.
%Each trajectory was stopped after moving down $10$ rows.
Complete details are in the appendix.

  This type of model (under constant field) has been applied to study electron motion
through insulators.\cite{achep01}
The zero-forcing case with elastic scattering
was studied analytically by Sinai,\cite{jlebo73,lbumo81} who showed that the trajectory of the
particle over long times converges to a Brownian random walk,
and that the expected direction of motion remains constant over time.
The evolution of the probability density can be shown to
converge to a Boltzmann transport equation,\cite{ggall69,cbold83}
and even has intuitive diffusive properties under a small,
constant external force.\cite{ncher08}
A review of approaches to the Lorentz gas was given by Spohn.\cite{hspoh80}
%In two dimensions, with a constant applied force, the periodic Lorentz
%gas is also known as the Galton board, which has an even longer history.
In the real-world case, the parabolic trajectories followed
by the particle make exact analysis difficult.
An analysis using a constant kinetic energy thermostat showed strong chaotic
properties, including fractal scaling of the probability
distributions for particle-scatterer impact.\cite{whoov89}
With elastic collisions, the kinetic energy of the particle must increase
linearly as the particle falls.  For this case, it has recently been shown that the particle velocity
grows with time as $t^{1/3}$, and that (analogous to the Gambler's ruin problem)
for large enough starting velocity the particle will
return to its initial height with probability 1.\cite{ncher09}
Our setup differs from these earlier studies because of the
presence of constant external force, inelastic collisions, and angular velocity.

\subsection{ Mode-Coupled Fermi-Pasta-Ulam Chains}

  To examine the time-course of energy redistribution between
harmonic modes of a crystal lattice, Fermi, Pasta, and Ulam (FPU) simulated
32 points moving in 1D with unit masses and coordinates,\cite{tdaux05}
$x_j, j=1, \ldots, N$.  This work uses periodic boundaries, so $x_0 = x_N$.
The potential energy function is,
\begin{align}
U(x) &= \sum_{j=0}^{N-1} V(x_{j+1} - x_j) \\
V(r) &= r^2/2 + \alpha r^3/3
.
\end{align}

  They discovered that for small anharmonic terms, energy did not seem to exchange,
but only to oscillate regularly between harmonic modes.
%Chirkov showed analytically that for $\alpha_3=0$ there are critical values of $\alpha_4$
%above which mixing can be achieved.\cite{}
Recent, much longer, simulations and theory have shown that
systems with small $\alpha$ do, in fact, equilibrate
but on an enormously long time-scale on the order of $\alpha^{-8}$.\cite{monor15}
This phenomenon has been explained as due to the nearness of
$V(r)$ with the potential for the Toda lattice, $e^{r} - r - 1$,
which is exactly integrable.

  To simulate this system numerically, we began by deriving
a symplectic, volume-preserving dynamical integration scheme
based on the Lagrangian,
\begin{equation}
L(x, \dot x) = \sum_j \frac{\dot x_j^2}{2} - V(x).
\end{equation}
Following the procedure of Marsden,\cite{jmars01}
we make the substitutions, $\dot x \to (x^{(t)} - x^{(t-1)})/\Delta t $,
$x \to x^{(t)}$, to construct a discrete action functional,
\begin{align}
A[\{x\}] &= \sum_{t=1}^T L_d(x^{(t)}, x^{(t-1)}) \\
L_d(x', x) &= \sum_j \frac{(x'_j - x_j)^2}{2 \Delta t} - \Delta t V(x')
.\label{e:Ld}
\end{align}

  Requiring stationary action yields,
\begin{equation}
\pd{L_d(x',x)}{x'} + \pd{L_d(x'',x')}{x'} = 0
,
\end{equation}
which translates to the Str\"{o}mer-Verlet scheme,
\begin{equation}
\frac{x'' - x'}{\Delta t} + \frac{x - x'}{\Delta t}
+ \Delta t \pd{V(x')}{x'} = 0
.\label{e:verlet}
\end{equation}
%This equation can be made exactly reversible if finite precision arithmetic
%is used to numerically solve for either $p' = (x'' - x') m/ \Delta t$ given $p = (x'-x) m/\Delta t$
%(in the forward direction) or for $-p$ given $-p'$ (in the reverse).
%Note that multiplication by $\Delta t$ effectively increases the precision in the order
%$m \delta x \sim \Delta t \delta p \sim \Delta t^2 \delta V'$.

  Rather than adding a thermostat to the coordinate-space integrator of Eq.~\ref{e:verlet},
we chose to cast the time integration in Fourier space,
\begin{equation}
X_k = \sum_{j=1}^N u^{-jk} x_j, \quad u \equiv e^{2\pi i/N}, \quad \omega_k^2 \equiv |u^k-1|^2 \label{e:FT}
.
\end{equation}
The versatility of the Lagrangian approach is exploited by
re-writing Eq.~\ref{e:Ld} in these new coordinates ($P_k \equiv (X_k' - X_k)/\Delta t$),
\begin{align}
L_d(X',X) %&= \Delta t  \sum_{j=0}^{N-1} \left[ \frac{(q_j'-q_j)^2}{2 \Delta t^2}  - (q_{j+1}' - q_j')^2/2 \right] - \Delta t V_c(q') \\
 &= \frac{\Delta t}{2 N} \sum_{k=0}^{N-1} \left[ |P_k|^2  - \omega_k^2 |X_k'|^2 \right] - \Delta t V_c(X')
 .
\end{align}
Here, $V_c$ represents the cubic potential terms.
According to MaxTrans, the exponent of the transition probability should be,
\begin{equation}
-\frac{\Delta t}{2\sigma^2} \left|\pdel{A[X(t)]}{X(t)}\right|^2 + \frac{\beta}{2} dX(t) \cdot \pdel{A[X(t)]}{X(t)}
.\label{e:FPUexp}
\end{equation}
Note how Eq.~\ref{e:en} behaves under a change in coordinate systems.
It simply amounts to transforming the deviations, $\sigma$, via the Jacobian, $|dx/dX|$.
The thermostatted equations of motion for $X_k$ can be read off from the mean and variance found
by factoring Eq.~\ref{e:FPUexp}
\begin{equation}
-\left(\pdel{A[X_k(t)]}{X_k(t)}\right)^* \Delta t = -\frac{\beta_k\sigma_k^2}{2} dX_k(t) + \sigma_k d\mathcal W_k
.
\end{equation}

  This equation of motion must be interpreted as applying to only $N$ degrees of freedom.
Since $X_k = X_{-k}^*$, the unique discrete Fourier variables are
Re[$X_0, X_1, \ldots, X_{\lceil (N-1)/2\rceil}$]
and Im[$X_1, \ldots, X_{\lfloor (N-1)/2\rfloor}$].
%The complex Wiener process $d\mathcal W_k$ requires
%$\dW_{-k} = \dW_k^*$, and should therefore have
%variance $\avg{|d\mathcal W_k|^2} = dt$.
%We accomplished this by rescaling $\sigma \to \sigma/\sqrt{2 dt}$
%and choosing independent standard normal variables for $\dW_k$.

  Energy exchange processes can be monitored by examining various
decompositions of the energy change (right side of Eq.~\ref{e:FPUexp}),
\begin{equation}
dE = dX^* \left(\frac{dP}{dt} - F_\text{harm} - F_\text{cubic} \right)
.
\end{equation}
By decomposing the total force as $dP/dt = F_\text{harm} + F_\text{cubic} + F_\text{lang}$,
we get
\begin{equation}
dE = dX^* F_\text{lang} = dX^* \left(\frac{dP}{dt} - F_\text{harm}\right) - dX^* F_\text{cubic} \\
.
\end{equation}
Of course, the first term on the right side just integrates to the sum
of energies in the unperturbed harmonic modes, $E_k(t) - E_k(0)$, where
\begin{equation}
E_k+E_{-k} = (|P_k|^2 + \omega_k^2 |X_k|^2)/2N \label{e:enk}
.
\end{equation}
We could separate the $\pm k$ parts by choosing a phase arbitrarily.
From this point of view, the time-derivative of $E_k$ (the energy in each mode), represents the
flux of energy from both the anharmonic system and the Langevin thermostat.

  To filter out noise coming from the Langevin thermostat,
we define the `heat flux' into mode $k$ to be
the integral of the second term, $dQ_k = dX^* F_\text{cubic}$.
It was computed numerically as the difference between the harmonic
oscillator's energy change and the energy added from
the Langevin thermostat.
Comparing to Eq.~\ref{e:enk} shows $E_k(t) - E_k(0) = Q_k + Q_k^\text{Langevin}$.
All heat flows, $Q$, thus come directly
from energy exchange through anharmonicity.
For individual modes that are coupled to a `hot' reservoir,
we will accordingly observe heat flow out of that
mode into anharmonic degrees of freedom.
No heat flow between modes is possible when $\alpha = 0$.
This was verified numerically to test our implementation.

  Fig.~\ref{f:sys}b shows example trajectories of energy flow from
mode $k=7$ to $k=3$ in the FPU system at $\alpha = 0.1$.
%The energy for mode $k$ was computed using the usual expression,
%in terms of the discrete Fourier modes (Eq.~\ref{e:FT}).
To provide a steady-state with energy flow, both modes $k=\pm 3$ and $k=\pm 7$ were
coupled to Langevin thermostats with $\beta_3 = 1 + \Delta\beta$ and $\beta_7 = 1 - \Delta\beta$,
respectively, and $\sigma_3 = \sigma_7 = 0.1$.  All other modes remained
un-thermostatted (equivalent to setting $\sigma^2_k = 0$).

  Distributions of $Q = Q_3 - Q_7$ presented here were calculated at time 1638.4
from a Verlet integration scheme with timestep $0.01$.
They include an initial transient of approximately 100 time units
because the initial conditions were chosen from the canonical distribution for
the harmonic system ($\alpha = 0$) at uniform temperature $\beta = 1$.

  Fig.~\ref{f:sys}c,d shows that the flow is a nonlinear function of the applied force
($\vec E$ for the Lorentz gas or $\Delta \beta$ for the FPU lattice).
This is especially apparent at large values of the forcing, where the probability
distributions over flow ($J$ for the Lorentz gas or $Q$ for the FPU lattice)
are markedly non-Gaussian.
These flows are totals, integrated over the first 10 rows for the Lorentz system
or the first 1638.4 time units for the FPU system.  Because of this they include
part of the initial transient as it relaxes to a conducting steady-state.

\section{ Discussion and Comparison}

%  This section provides further background and
%highlights the connections between theories
%by presenting short derivations of alternative approaches
%to the present problem.

  Often, applied literature provides specialized fluctuation-dissipation
or fluctuation theorems that give little hint as to how they may be generalized
or extended.  In fact, the original derivations allow quite a bit of flexibility in defining
what forces and flows can enter, and can be put into a form very much resembling
our major results (Eqns.~\ref{e:fdt2},~\ref{e:fdt3}, and~\ref{e:maxtrans}).
We discuss these alternative viewpoints by standardizing the notation
and re-stating the theorems in terms of time-derivatives (flows)
rather than absolute positions.

\subsection{ Projector-Operator and Fluctuation-Dissipation Theorems}

  The projector-operator theory gives a rigorous, general equation of motion
for the probability distribution of coarse coordinates like the particle position or
the energy in each mode. %\cite{rzwan83}
The theory clearly indicates where closure relations
are required.  This section shows how the simplest closure
relations with Gaussian noise can be derived by analogy to Gaussian processes.
The result provides time-dependent Green-Kubo relations applicable at
nonzero driving force.  They are linear because they predict only the slope of the
flow {\em vs.} force curve.\cite{droge11a}

%  The response theory makes predictions about how average flows
%respond to external driving.
%While the theory is commonly used to understand how correlated,
%microscopic fluctuations determine the response of the average,
%it also makes predictions for how the entire
%probability distribution responds to external driving as well.
%It is therefore correct to say that fluctuation theorems state
%facts about the symmetries of the response theory.

    An accessible derivation of the projector-operator theory was given by Nordholm
and Zwanzig\cite{snord75} with the result,
\begin{align}
&\pd{}{t} \P f(t, \Gamma) = - \P i\mathcal L \P f(t,\Gamma) \notag \\
&+ \int_0^t \; ds \P i\mathcal L
e^{-is(1-\P) \mathcal L} (1-\P) i\mathcal L \P f(t-s, \Gamma) \notag \\
&-\P i\mathcal L e^{-it(1-\P) \mathcal L} (1-\P) f(0,\Gamma). \label{e:proj}
\end{align}
The operator, $\P$, projects the phase-space probability density, $f(t,\Gamma)$,
onto a subspace of the full phase space, $\Omega = \{\Gamma\}$.
The Liouville operator is defined in terms of the Poisson brackets, $i\mathcal L J = \{J, H\} = dJ/dt$.
There is no difficulty interpreting this subspace as an arbitrary manifold lying inside $\Omega$.
For any point, $\Gamma$, we can define the projected point
on the manifold as $\phi(\Gamma)$ so that
\begin{equation}
\P f(\Gamma) = \int_\Omega d\Gamma' \; \delta(\Gamma - \phi(\Gamma')) f(\Gamma')
.
\end{equation}

  The projected equation of motion (Eq.~\ref{e:proj}) implicitly defines the
probability distribution of transition events, $g = \phi(\Gamma) \to \phi(\Gamma')$.
It is trivial to re-cast it in this way, since $dt \pd{}{t} \P f(t,\Gamma) =
\int dg \; \Pr{\Gamma | \Gamma', dt} f(t', \Gamma') - f(t',\Gamma)$
as the timestep, $dt \to 0$.

  The push-forward operation, $i\mathcal L \P f(t,\Gamma)$, in both of the first two
terms refers explicitly to this transition.
The three parts of the equation of motion on the manifold (Eq.~\ref{e:proj})
have the interpretation of
{\em i}) the deterministic transition ($\P i\mathcal L \P$) for points on the manifold,
{\em ii}) the memory function describing the predictable,
but delayed effect due to earlier transitions, $(1-\P) i\mathcal L \P f(t-s, \Gamma)$,
and {\em iii}) the `random' noise part due to initial conditions not on the manifold.

  Because they are off the manifold, the equation of motion makes
it clear that closure relations are required for describing
parts ({\em ii}) and ({\em iii}).  Specific choices for those closures
form the starting points for mode coupling theory\cite{snord75}
and nonlinear fluctuating hydrodynamics.\cite{gmaze06}
That theory has also been applied to dynamics in large-FPU chains.\cite{cmend13}

  In practice, most applications of the theory have used linear closure relations, which
give rise to linear transport equations.\cite{rzwan65}
Nonlinear closures are most easily understood
by comparison to the linear theory.\cite{rkubo66}
It has been pointed out\cite{ejayn80} that the principle results of the linear theory
are identical to linear regression.

  The linear regression case has been treated in a very general way in the Gaussian process
literature.  The critical assumptions are that the random noise obeys Gaussian statistics
and that the coefficients of the memory function depend only on time,
not on the process history.
The equations below relate to the two systems considered here by
replacing $g$ with the horizontal motion of the disk, $dJ$,
or the heat transfer, $dQ$, over a small amount of time.
The regression equations can be summarized by the assumption,\cite{crasm06}
\begin{equation}
\Pr{\{g(t_i)\}_0^n} = GP[m(t_i), k(t_i,t_j)] \label{e:GP}
\end{equation}
which implies the following generating process,
\begin{align}
g(t_n) &= m(t_n) + \sum_{j,k = 1}^{n} k(t_{n}, t_{n-j}) k^{-1}_{n-j; n-k} \notag \\
& \times (g(t_{n-k})-m(t_{n-k})) + \sigma_n R_n \label{e:fdt}
\end{align}
\begin{equation}
\sigma^2_n = k(t_n,t_n) - \sum_{j,k = 1}^{n} k(t_{n}, t_{n-j}) k^{-1}_{n-j; n-k} k(t_{n}, t_{n-k}) \label{e:sigma}
\end{equation}
Here, $GP$ denotes a Gaussian process, which is a multivariate normal distribution
with mean $m(t)$ and variance-covariance matrix $k(t,t')$.
Eq.~\ref{e:fdt} states the applicable fluctuation-dissipation theorem -- namely that
the probability of $g$ at the next step has a Gaussian distribution with a mean
that is linear in the random increments, $g(t_k) - m(t_k)$,
and a variance, $\sigma_n^2$, that is reduced by knowledge of the process history.
The variable $R_n$ is an independent sample from the standard
normal distribution.

  This closure is demonstrated by noting the terms in Eq.~\ref{e:fdt}
correspond 1:1 with those of Eq.~\ref{e:proj}.
Brownian motion theory is recovered when $g$ is taken to be the
momentum.  Then $m(t)$ is the drift velocity and $k(t,t) = k_BT/m$
near equilibrium.  Its time-derivative describes Langevin dynamics,
where $g_n$ is the momentum update, $p_n - p_{n-1}$.

  From these identifications, a little algebra shows that
applying a single external force at time $t_0$ will not
only directly shift $g_0 \to g_0 + F^\text{ext}_0$,
but will also accumulate a net effect at later times, $t_k$ of,
\begin{align}
\delta g_{k\gets 0} &= \begin{bmatrix} k(t_k, t_I)
\end{bmatrix}^T k^{-1}_{IJ} \begin{bmatrix}
\delta g_0 \\ \delta g_{1\gets 0} \\ \vdots \\ g_{k-1\gets 0}
\end{bmatrix}
\intertext{The recursion is solved by}
\delta g_{k\gets 0} &= \frac{k(t_k, t_0)}{k(t_0, t_0)} F^\text{ext}_0
.
\end{align}

  In comparison with our main result (Eq.~\ref{e:fdt2}), this externally forced process could have
been derived extremely easily by adding an
exponential bias to the basic Gaussian process (Eq.~\ref{e:GP}),
\begin{align}
\Pr{\vec g | F^\text{ext}} \propto \exp \Big\{
  &- (\vec g - \vec m)^T k^{-1} (\vec g-\vec m)/2  \notag \\
  &+ (\vec g - \vec m)^T (F^\text{ext}/\mathrm{diag}(k))
\Big\}
. \label{e:bias}
\end{align}
This is the revised Onsager-Machlup action functional approach.\cite{lonsa53,lbert15}

  The linear transport theory can be re-derived in a simplified way as a Gaussian
process.  The technical content of the celebrated fluctuation-dissipation
theorem in this case is a statement of how dissipation of an external force,
$\sum_{j<n} k(t_n,t_{n-j})$,
is related to fluctuations of the current, $k(t_i,t_j) = \avg{\delta g_i \delta g_j}$.

  We can see that this line of attack applies to time-dependent processes,
but Gaussian processes do not make it clear how to extend the theory into the
nonlinear regime.
The major contribution of Sec.~\ref{s:maxtrans} was to
replace the fitting ansatz of Eq.~\ref{e:GP}
with a single-step fitting ansatz at time $t_i$.
This frees $m_i$ and $k_{ij}$ to be
arbitrary nonlinear functions of the history, $\{g\}_0^{i-1}$.
However, it sacrifices knowledge of the steady-state
properties.

% and using Eqns.~\ref{e:fdt} and~\ref{e:sigma}.
%Also, for the two example problems given here, Eq~\ref{e:fdt}
%does not help much in prediction.
%This is because the external force must be identified by its effect on the instantaneous average,
%$F(t) = \delta A(t)$.
%This would be known for each individual time-step of Hamiltonian dynamics,
%but finding the correct long-time relationship is a difficult task, even for
%for the Galton board\cite{achep01,ncher08}
%and FPU heat models.\cite{jeckm99,hspoh14,monor15}

\subsection{ Fluctuation Theorems and Chaotic Hypothesis}

  Fluctuation theorems address the probabilities of transitions even more directly.
Specifically, they transform symmetries of the dynamical equations
into symmetries of integrated quantities such as work and heat.
They have a history stretching back to Callen and Welton,\cite{hcall52}
who proved a fluctuation theorem showing the odds of heat, $Q$, flowing from
cold to hot vs. the reverse process was proportional to $\exp(Q\Delta \beta)$.

  Since the literature on fluctuation theorems is large, I provide here only
a few results.  The first fluctuation theorems about
atomistic trajectories were developed
by several groups,\cite{devan02,ggall08} who proved theorems of the form,
\begin{equation}
\frac{1}{t} \log \frac{\Pr{g = \sigma(t,\Gamma)/t | x_0 \to x_n}}
{\Pr{ g = -\sigma(t,\Gamma)/t | x_n \to x_0}} \asymp g
\label{e:FT}
\end{equation}
The symbol, $\asymp$, means asymptotic convergence with large time, $t$.
The conditioning on coordinates, $x_0$ or $x_n$, indicates
whether the trajectory is initiated from a starting or ending point.
For the transient fluctuation theorem, the microscopic entropy
production is identified with the time-integral
over a trajectory of length $t$ starting from $\Gamma(0) = \Gamma$,
\begin{equation}
\sigma(t, \Gamma) = \int_0^t dt \; D(\Gamma(t))
,
\end{equation}
where $D(\Gamma)$ is the ratio of phase space volume between
the last and next time-step, $| i\mathcal L (d\Gamma) | / |d\Gamma|$.
Because of the dependence on the starting/ending point,
there are differences in the relations and proofs depending on whether the starting
states are fixed or chosen at random from an SRB measure (steady-state),
whose existence and uniqueness requires additional assumptions.\cite{ggall96}

  In the case where a dynamical system can be modeled as a finite-state
Markov process, a new version of the fluctuation theorem (Eq.~\ref{e:FT})
can be shown,\cite{jlebo99,gcroo00}
where $\sigma$ ($W$ of Ref.~\citenum{jlebo99}) is the log-ratio of
forward to reverse transition probabilities over $t = n$ steps of the Markov process,
\begin{equation}
\sigma(n,\Gamma) = \sum_{i=0}^{n-1} \log \frac{\Pr{\Gamma_{i+1} | \Gamma_i}}{\Pr{\Gamma_i | \Gamma_{i+1}}}
.\label{e:smark}
\end{equation}
Although they apply in different cases, the two fluctuation relations
essentially express the same measure of
irreversibility, since the probability of a transition scales inversely with the starting volume,
$\Pr{\Gamma_{i+1} | \Gamma_i} \propto 1/|d\Gamma_i|$.\footnote{
To make this statement rigorous requires comparing the number of transitions that can
be made out of state $\Gamma_{i}$ vs. those out of state $\Gamma_{i+1}$.}

  Since the log-ratio of transition probabilities are often 
related to work and entropy production, the fluctuation theorems
can make quantitative statements about energy exchange
during transitions of a dynamical system.
Although Eq.~\ref{e:FT} appears to be a special case of Eq.~\ref{e:maxtrans},
Eq.~\ref{e:FT} has been proven to hold under more general conditions.
Its relation to symmetry provides it with a unique
status in that it is closer to a dynamical
law than a statistical one.

  It is also possible to specialize $\sigma$
for describing transition probabilities of other coarse variables.\cite{devan02,ggall95}
This interpretation often refers to $\sigma$ as a dissipation function,
or, in reference to the Onsager-Machlup theory, as an action function.
Our alternative derivation of $\sigma$ as a generalized entropy
in section~\ref{s:maxtrans} provides a more canonical explanation of
this connection between $\sigma$ and force/flux relations.

\begin{figure*}
{\centering
\includegraphics[width=7cm]{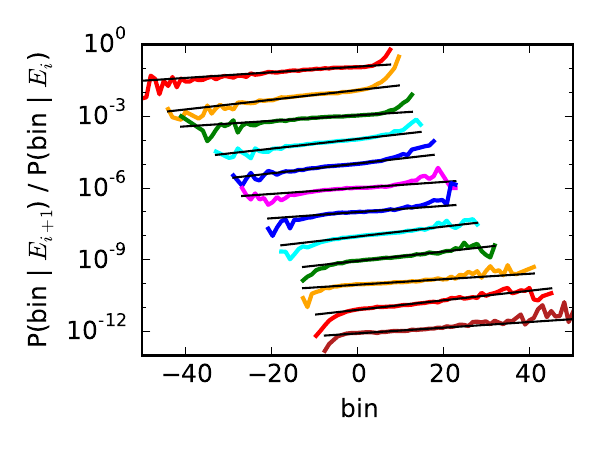}
\includegraphics[width=7cm]{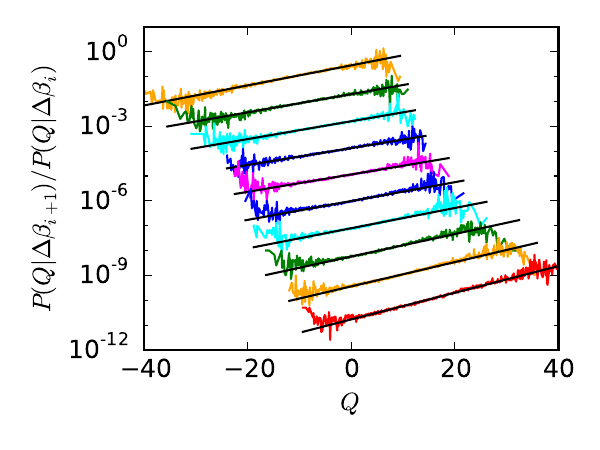} }\vspace{-2em}
\caption{Log-probability ratios for successive values of the applied force.
Forces increase left-to-right, with each line shifted down by a factor of 10 for
visual separation.  Black lines show linear fits to these log-ratios.
The left panel shows data from the Galton board, while the
right shows heat conduction in the FPU lattice.}\label{f:ft}
\end{figure*}

  Postulates like Eq.~\ref{e:pred} have been
hinted at before in connection with Onsager-Machlup theory\cite{hfors08},
but their generality and use for deriving fluctuation theorems
has not been widely appreciated.
A similar theory of Jaynes, named maximum caliber (MaxCal),\cite{ejayn80}
applied maximum entropy to the set of all trajectories.
Unfortunately, that theory does not maintain causality,\cite{droge12}
since forces applied in the future influence the entire history of the process.
This shortcoming has not caused identifiable problems where the
theory has been applied,\cite{gstoc08,spres11,kdill15}
since maximum caliber and MaxTrans make the same probability
assignments when there is time translation invariance of the forces.
In these situations, non-causality only creates problems when applying the
fluctuation theorems.  The situation can be summed up by noting
that path counting gives non-causal weights to trajectories,
whereas the transition probabilities are always causal.

  Both Jaynes\cite{ejayn85} and Haken\cite{hhake85,hhake93} investigated maximizing
the entropy of the transition distribution as a restriction of MaxCal.
Those early works chose steady-state averages as constraints, rather than
the instantaneous flows as done here.
Since the Lagrange multipliers had to be identified through the Fokker-Planck equation,
that choice hid the connection to the equilibrium, Boltzmann/Gibbs distribution.
Had they made the latter choice, they would have immediately discovered
our kinetic partition function, $\Z$, along with its attendant force-flux
relationships and FDTs.

\begin{figure*}
{\centering
\includegraphics[width=7cm]{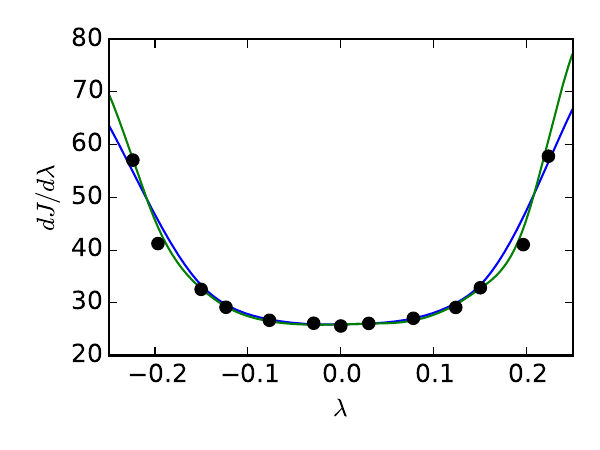}
\includegraphics[width=7cm]{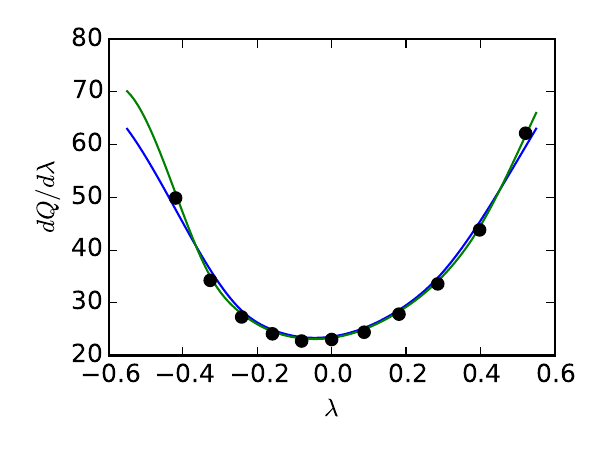} }\vspace{-2em}
\caption{Comparison between asymptotically equivalent expressions for the conductivity coefficient.
The lines show different numerical estimates of derivatives of the flux vs. force curves ($d\avg{J}/d\lambda$) computed from alternate finite difference methods, while the variance of the flux ($\avg{\delta J^2|\lambda}$) is shown as individual points.  The blue line uses a 3$^\text{rd}$ order finite difference interpolation of $J(\lambda)$ (Fig.~\ref{f:sys}c,d) and the green line uses a 7$^\text{th}$ order interpolation from the same data.
The left panel shows data from the Galton board, while the right shows heat conduction in
the FPU lattice.}\label{f:slope}
\end{figure*}

  Figure~\ref{f:slope} compares properties {\em i} and {\em ii} (listed at the
end of Sec.~\ref{s:maxtrans}) by plotting $d\avg{J|\lambda}/d\lambda$
(lines) from the force-flux curves of Fig.~\ref{f:sys} against
the fluctuations $\avg{\delta J^2}_\lambda$ (shown as points).
The two smooth lines on each plot come from the derivatives obtained from
3$^\text{rd}$ and 7$^\text{th}$ order numerical differencing schemes.
Differences between these lines are a measure of the uncertainty in
collected force {\em vs.} flux data.
The correspondence, $\lambda = \lambda(E)$ or $\lambda(\Delta\beta)$, was
made by the fits to the slopes
in Fig.~\ref{f:ft}.  Linear response predicts that $\avg{J}$ will be proportional to
$\lambda$, $\lambda$ proportional to the force, and hence $\avg{\delta J^2}$ will be constant.
While $\avg{\delta J^2}$ is almost constant near the origin for the Lorentz gas,
linear response clearly does not hold in the FPU system.
Nevertheless, our numerical data show the validity of Eq.~\ref{e:maxtrans}
well out of the linear response regime.
Fig.~\ref{f:ft} shows method {\em i} and Fig.~\ref{f:slope} shows method {\em ii}.
The fluctuations of the current still correspond to the slope of the flux-force curve away
from the region of ``constant'' $\avg{\delta J^2}$ around $\lambda = 0$.

  The canonical form of Eq.~\ref{e:langevin} codifies flux/force relations in a coordinate-independent way
using a maximum entropy structure.
Because of this, it provides facile derivations for both time-dependent
Green-Kubo response theories and the fluctuation theorems
when the postulate of Eq.~\ref{e:pred} holds.
Its shortcoming is that it does not directly predict the relationship
between the forces and flows, e.g. $\beta$ and $\avg{dE}$.
This, however, is exactly the well-known problem of determining
the equations of state in equilibrium statistical mechanics.

\section{ Conclusions}

  We have constructed a probability distribution over values of flows that occur
during transitions using maximum entropy.
%the tools of maximum entropy and Bayesian inference.
%Three complementary approaches to describing general force/flux relationships in
%nonequilibrium statistical mechanics were compared.
It was shown that the nonlinear, maximum entropy structure
is a general consequence of an unknown, and hence subjectively random, environment.
For atomistic problems, a general prescription exists to derive a generalized
Langevin equation from an action functional.
For coarse-grained properties, Eq.~\ref{e:pred} just
states the maximum entropy postulate and
must be experimentally verified wherever it is used.
Numerical simulationsof driven diffusion in 2D gas models 
and heat diffusion between modes of a crystal
supported it.
We expect the postulate to be more easily satisfied
as the number of transitions grows -- even for transient, driven processes.
The success of the linear response methods
and of the fluctuation theorems both rest on exploiting the existence of this
structure within the transition distribution.

  This theory has been directly related to two complementary methods of attack.
The Green-Kubo relations provide quick estimates for
the conductivities, even in transient steady-states.
This work showed the connection by focusing on the transition
quantities in Eq.~\ref{e:fdt}, which have the same structure as Eq.~\ref{e:fdt2}
and~\ref{e:fdt3}.
Fluctuation theorems provide connections with distributions of work and notions of irreversibility.
We have shown the forward fluctuation relation of Eq.~\ref{e:maxtrans}
derives Eq.~\ref{e:FT} when the flux is related to number of transitions.
Transition distributions provide an entry point to these theories,
but can also be applied in their own right as a maximum entropy structure.
Although it derives mathematically identical results, it is logically
distinct from large deviation theories
because maximum entropy infers transition distributions even when the underlying
microscopic noise process is unknown.
It is also distinct from Gaussian processes or near-equilibrium theories that use the
idea of local entropies because it focuses specifically on
one-step transitions.\cite{lonsa53,iprig65,hspoh78,ejayn80}
All general approaches are structural in the sense that finding analytical expressions for
long-time force/flux relationships is a difficult task, even for
for the Lorentz gas\cite{achep01,ncher08}
and FPU heat models.\cite{jeckm99,hspoh14,monor15}
Our line of reasoning avoids that problem, and instead parallels the reasoning
used to derive equilibrium equations of state.  Its most common criticism
-- that certain analytic properties of the microscopic distribution must be proven --
directly parallels the integrability objection to the ergodic hypothesis.

\section*{ Acknowledgments}

  I thank the anonymous reviewers for
suggesting improvements to the presentation of this work.
This work was supported by the USF Research Foundation and
NSF MRI CHE-1531590.

%\nocite{TitlesOn}
\bibliographystyle{unsrt}
%\bibliography{stat}

\appendix{}

%\begin{equation}
%dp_j = (V'(q_{j+1}-q_j) - V'(q_j - q_{j-1})) dt + \operatorname{Re}\left[
%\sum_{k=0}^{\lceil(N-1)/2\rceil} u^{jk} \left( - \frac{\beta_k\sigma_k^2}{2} P_k dt + \sigma_k \dW_k \right)
%\right] \label{e:therm}
%.
%\end{equation}

%Note this also enforces the necessary restriction $\sigma_k,\beta_k = \sigma_{-k},\beta_{-k}$.
%\begin{align}
%E_k = \frac{1}{2N} (|P_k|^2 + \omega_k^2 |Q_k|^2), \label{e:enk} \\
%.
%\end{align}

\end{document}